\documentclass[conference,a4paper]{IEEEtran}
\usepackage{amsmath,amsfonts,amssymb}
\usepackage[center]{caption}
\usepackage{graphicx}
\usepackage[center]{subfigure}
\usepackage{multirow,bigdelim,longtable}
\usepackage{amsthm}
\usepackage{hhline}
\usepackage{tikz,pgfplots}
\usetikzlibrary{intersections,calc,patterns,arrows}
\usepackage{float}
\usepackage{flushend}
\restylefloat{table}
\theoremstyle{definition}
\newtheorem{thm}{Theorem}
\newtheorem{lem}{Lemma}

\newtheorem{cor}{Corollary}
\newcommand{\squeezeup}{\vspace{-2.5mm}}
\IEEEoverridecommandlockouts

\begin{document}

\sloppy

%% Paper Title
%% You can use linebreaks \\ within to get better formatting as
%% desired. 
\title{Generalized Diversity-Multiplexing Tradeoff of Half-Duplex Relay Networks}

%% Author names and affiliations:
%%
%% Avoiding spaces at the end of the author lines is not a problem with
%% conference papers because we don't use \thanks or \IEEEmembership.
%%
%% For several authors with only one affiliation:
%%
% \author{
%   \IEEEauthorblockN{Hui-Ting Chang and Stefan M.~Moser}
%   \IEEEauthorblockA{Department of Electrical and Computer Engineering\\
%     National Chiao Tung University (NCTU)\\
%     Hsinchu, Taiwan\\
%     Email: \{email-of-hui-ting,email-of-stefan\}@ieee.org} 
% }
%%
%% For up to three affiliations:
%%
\author{
  \IEEEauthorblockN{Ritesh Kolte and Ayfer \"{O}zg\"{u}r}
  \IEEEauthorblockA{Stanford University, California 94305 USA\\
    Email: \{rkolte,aozgur\}@stanford.edu} 
\thanks{The work of R. Kolte and A. \"{O}zg\"{u}r was supported in part by a Stanford Graduate Fellowship and NSF CAREER award 1254786 respectively.}    
}
\maketitle

\squeezeup\squeezeup
\begin{abstract}
Diversity-multiplexing trade-off has been studied extensively to quantify the benefits of different relaying strategies in terms of error and rate performance. However, even in the case of a single half-duplex relay, which seems fully characterized, implications are not clear.  When all channels in the system are assumed to be independent and identically fading, a fixed schedule where the relay listens half of the total duration for communication and transmits  the second half combined with quantize-map-and-forward relaying (static QMF) is known to achieve the full-duplex performance \cite{PawAveTse}. However, when there is no direct link between the source and the destination, a dynamic decode-and-forward (DDF) strategy is needed \cite{GunKhoGolPoo}. It is not clear which one of these two conclusions would carry to a less idealized setup, where the direct link can be neither as strong as the other links nor fully non-existent.

In this paper, we provide a generalized diversity-multiplexing trade-off for the half-duplex relay channel which accounts for different channel strengths and recovers the two earlier results as two special cases. We show that these two strategies are sufficient to achieve the diversity-multiplexing trade-off across all channel configurations, by characterizing the best achievable trade-off when channel state information (CSI) is only available at the receivers (CSIR). However, for general relay networks we show that a generalization of these two schemes through a dynamic QMF strategy is needed to achieve optimal performance.
\end{abstract}

\section{Introduction}
The diversity-multiplexing tradeoff (DMT) \cite{ZheTse} captures the fundamental tension between reliability and rate in fading channels. It has been also used to study the value of relays in providing reliability and/or rate gains~\cite{LanTseWor,SenErkAaz,AzaGamSch}. The two critical issues that complicate the problem in relay networks is who knows what channel state and whether nodes can listen and transmit at the same time (i.e. half or full-duplex).  The full-duplex case is fully characterized as a consequence of the recent constant gap approximation of the capacity of wireless networks in \cite{AveDigTse}. For any statistics of the channel fadings and any network topology, it suffices for the  relays to quantize-map-and-forward (QMF) their observations to achieve the optimal DMT trade-off. This strategy only requires the relays to know their incoming channel states (CSIR) as opposed to earlier strategies which require global CSI \cite{YukErk}. Typically CSIR is much easier to obtain than transmit CSI (CSIT) in a fading environment giving rise to an outage setting. 
 
In current wireless systems, however, nodes operate in a half-duplex mode, i.e., they can not simultaneously transmit and receive signals on the same frequency band. Designing DMT optimal strategies for half-duplex networks is more challenging as it also involves an optimization over the listen and transmit times for the relays. In a fading environment where CSIT is unavailable at the nodes, such a listen-transmit schedule needs to be either fixed or depend only on local CSIR. Characterizing the DMT of general half-duplex relay networks remains an open problem, and even in the special cases where the DMT has been characterized implications are not clear.

\begin{figure}
\centering
\subfigure[]{
\begin{tikzpicture}[scale=1.2]
\node (source) at (0,0) [circle,draw,fill,inner sep=0pt,minimum size=2mm,label=left:$S$] {};
\node (relay) at (1,0.5) [circle,draw,fill,inner sep=0pt,minimum size=2mm,label=above:$R$] {};
\node (dest) at (2,0) [circle,draw,fill,inner sep=0pt,minimum size=2mm,label=right:$D$] {};
\path[draw,->,>=stealth'] (source)  -- (relay);
\path[draw,->,>=stealth'] (relay)  -- (dest);
\path[draw,->,>=stealth'] (source)  -- (dest);
\end{tikzpicture}
\label{subfig:relay}}
\subfigure[]{
\begin{tikzpicture}[scale=1.2]
\node (source) at (0,0) [circle,draw,fill,inner sep=0pt,minimum size=2mm,label=left:$S$] {};
\node (relay) at (1,0) [circle,draw,fill,inner sep=0pt,minimum size=2mm,label=above:$R$] {};
\node (dest) at (2,0) [circle,draw,fill,inner sep=0pt,minimum size=2mm,label=right:$D$] {};
\path[draw,->,>=stealth'] (source)  -- (relay);
\path[draw,->,>=stealth'] (relay)  -- (dest);
\end{tikzpicture}
\label{subfig:line}}\squeezeup
\caption{(a) Single relay network, (b) Line relay network}\squeezeup\squeezeup
\end{figure}
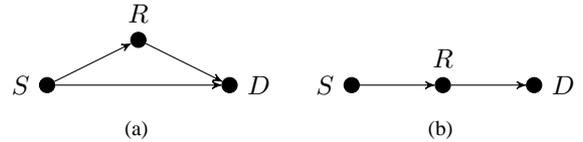

Consider the simplest case where the communication between a source and a destination is assisted by a single relay. Two setups for a single relay network have been considered and characterized in the literature:
\begin{itemize}
\item When all links (source-destination, source-relay, relay-destination) are independent and identically fading (see Figure~\ref{subfig:relay}), \cite{PawAveTse} shows that the optimal DMT is achieved by the QMF scheme with a \emph{fixed RX-TX} schedule for the half-duplex relay. Here, the relay listens half of the total duration for communication, then quantizes and maps its received signal to a random codeword and transmits it in the second half. We call this strategy static QMF in the sequel. The performance meets the full-duplex DMT.  
\item When there is no link between the source and the destination, the single relay channel of Figure~\ref{subfig:relay} reduces to the line topology in Figure~\ref{subfig:line}. In this case, \cite{GunKhoGolPoo} shows that the optimal DMT is achieved by a dynamic decode-and-forward (DDF) strategy at the relay. In DDF, the relay listens until it gathers enough mutual information to decode the transmitted message so its RX time is dynamically determined as a function of its CSIR and the targeted rate  \cite{AzaGamSch}. The optimal performance does not reach the full-duplex DMT.
\end{itemize}
In a practical setup, the source-destination link can be expected to be neither as strong as the relay links nor fully non-existent. The two results above address the two extreme cases. Given the difference in the nature of the optimal strategies and the optimal DMT trade-off in these two extremes, natural questions are: which of these conclusions apply to a general setup where channel strengths are arbitrary; do we need new strategies to achieve the optimal DMT in the general case? 

In this paper, we answer these questions by presenting a generalized diversity-multiplexing trade-off analysis for the single relay channel. Instead of considering a single limit for the average SNR's of different channels, we consider a family of limits where the SNR's of different links scale at different rates. This allows to capture  the impact of different channel strengths within the DMT framework. The generalization is similar in spirit to generalized degrees of freedom in \cite{EtkTseWan} or the operating regimes in \cite{OzgJohTseLev}. In the study of the DMT, the idea has been first introduced by \cite{NagPawTseNik}, where it is used to demonstrate the usefulness of relays for achieving cooperative multiplexing gain.

\subsection{Main Results}
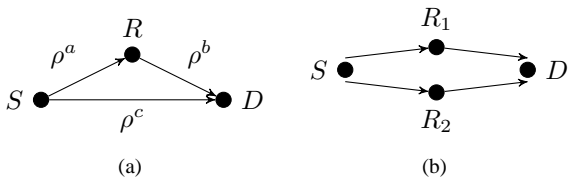
\begin{figure}
\centering
\subfigure[]{
\begin{tikzpicture}[scale=1.2]
\node (source) at (0,0) [circle,draw,fill,inner sep=0pt,minimum size=2mm,label=left:$S$] {};
\node (relay) at (1,0.5) [circle,draw,fill,inner sep=0pt,minimum size=2mm,label=above:$R$] {};
\node (dest) at (2,0) [circle,draw,fill,inner sep=0pt,minimum size=2mm,label=right:$D$] {};
\path[draw,->,>=stealth'] (source)  -- node[above left] {$\rho^a$} (relay);
\path[draw,->,>=stealth'] (relay)  -- node[above right] {$\rho^b$} (dest);
\path[draw,->,>=stealth'] (source)  -- node[below] {$\rho^c$} (dest);
\end{tikzpicture}
\label{subfig:abc_relay}}
\subfigure[]{
\begin{tikzpicture}[scale=1.2]
\node (source) at (0,0) [circle,draw,fill,inner sep=0pt,minimum size=2mm,label=left:$S$] {};
\node (relay1) at (1,0.25) [circle,draw,fill,inner sep=0pt,minimum size=2mm,label=above:$R_1$] {};
\node (relay2) at (1,-0.25) [circle,draw,fill,inner sep=0pt,minimum size=2mm,label=below:$R_2$] {};
\node (dest) at (2,0) [circle,draw,fill,inner sep=0pt,minimum size=2mm,label=right:$D$] {};
\node (out_source) at (0,0) [circle,inner sep=0pt,minimum size=3mm] {};
\node (out_dest) at (2,0) [circle,inner sep=0pt,minimum size=3mm] {};
\path[draw,->,>=stealth'] (out_source.north)  -- (relay1);
\path[draw,->,>=stealth'] (out_source.south)  -- (relay2);
\path[draw,->,>=stealth'] (relay1)  -- (out_dest.north);
\path[draw,->,>=stealth'] (relay2)  -- (out_dest.south);
\end{tikzpicture}
\label{subfig:parallel}}
\squeezeup\caption{(a) Generalized relay channel, (b) Parallel relay channel}\squeezeup\squeezeup
\end{figure}

\begin{figure}
\centering
\input{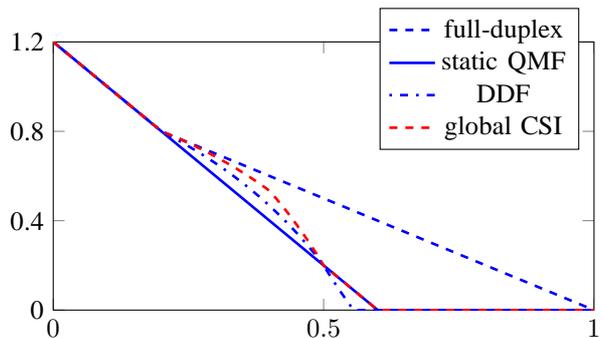}
\squeezeup\caption{DMT for $(a,b,c)=(1,1,0.2)$}\squeezeup\squeezeup
\label{fig:dmt_plot}
\end{figure}

\squeezeup Let $(a,b,c)$ be the exponential orders of the average SNR's of the source-relay (S-R), relay-destination (R-D) and source-destination (S-D) channels respectively and $r$ be the desired multiplexing rate. See Figure~\ref{subfig:abc_relay}. We show that:
\begin{itemize}
\item when $c\geq\min(a,b)$, i.e. when the S-D link is as strong as or stronger than one of the relay links, static QMF achieves the full-duplex DMT. The result of \cite{PawAveTse} corresponds to the special case $(a,b,c)=(1,1,1)$.
\item  when $c <\min(a,b)$, i.e. the S-D link is weaker than one of the relay links but $r\leq c$, then the full-duplex DMT can still be achieved by static QMF.
\end{itemize}
The remaining regime is when $c <\min(a,b)$ and $r>c$. To simplify the analysis, we concentrate on the case where $a=b=p$.  We show that:
\begin{itemize}
\item when $r\geq p/2$, static QMF is again DMT optimal. It does not achieve the full-duplex DMT in this case, but it does achieve the best DMT under the more optimistic assumption that the TX-RX schedule can be optimized based on the knowledge of all instantaneous channel realizations in the network (i.e. global CSI at the relay). The result implies that this additional CSI is not needed. The largest achievable multiplexing gain is given by $\frac{p+c}{2}$. 
\item when $c < r< p/2$, we show that DDF achieves the optimal DMT under local CSIR. In this case, global CSI can improve the DMT. To the best of our knowledge, this is the first upper bound on the DMT trade-off under local CSIR. The result of \cite{GunKhoGolPoo} corresponds to the special case $(a,b,c)=(1,1,0)$.
\end{itemize}
These conclusions are summarized in Figure~\ref{fig:dmt_plot}.\footnote{The fact that DDF is optimal for small multiplexing gains and QMF with a fixed schedule is optimal for large multiplexing gains is similarly the case when the relay is equipped with multiple antennas \cite{KarVar,PraVar}. However, the two settings and the resultant trade-offs are quite different. For example, here in the very low multiplexing gain regime, QMF with a fixed schedule also becomes optimal, or in the intermediate multiplexing gain regime DDF is optimal but cannot achieve the global CSI upper bound. This is not the case with multiple antennas.}

The above discussion shows that static QMF and DDF are sufficient to achieve the optimal trade-off in the single-relay channel. A natural
generalization of these two strategies is dynamic QMF where
the relay listens for a fraction of time determined by its CSIR that is not necessarily long enough to allow decoding of the transmitted message. The relay then quantizes, maps and forwards the received signal as in the original QMF \cite{AveDigTse}. \cite{OzgDig} shows that this generalization is necessary to achieve the optimal trade-off in networks with multiple relays, through the parallel relay channel in Figure~\ref{subfig:parallel}. However, it fails to characterize the optimal RX times for the relays and the corresponding optimal trade-off. We close this gap in the current paper.

%\begin{figure}
%\centering
%\includegraphics[scale=0.6]{dmt_fig}
%\caption{DMT under different protocols for the case $c<\frac{p}{2}$}
%\label{fig:dmt_plot}
%\end{figure}

\section{System Model}\label{sec:model}

\squeezeup We consider the two configurations depicted in Figure \ref{subfig:abc_relay} and \ref{subfig:parallel} in Sections \ref{sec:relay} and \ref{sec:parallel} respectively. In the first case, the source transmission is broadcasted to the relay and destination, while the source and relay transmissions superpose at the destination. The relay is half-duplex. All channels are assumed to be flat-fading, i.e. the channel gains for every link are i.i.d. circularly-symmetric complex Gaussian random variables $\mathcal{CN}(0,1)$. Channel gain between nodes $A$ and $B$ is denoted by $h_{ab}$. We assume quasi-static
fading, i.e. the channel gains remain constant over the duration of the codeword and change independently from one codeword to another. Local channel realizations are known at the receivers (CSIR) but not at the transmitters (no CSIT). The classical DMT formulation assumes that all channels in the network operate at the same average SNR. However, the average SNR's of different channels can be significantly different due to path-loss and shadowing. 
Following \cite{NagPawTseNik}, we assume that the average SNR's of the S-R, R-D and S-D  links are $\rho^a$, $\rho^b$ and $\rho^c$ respectively for $a,b,c\geq 0$. A sequence of codes $\mathcal{C}(\rho^a, \rho^b, \rho^c)$ indexed by $\rho^a$, $\rho^b$ and $\rho^c$ with rate $R(\rho^a, \rho^b, \rho^c)$ and average error probability $P_e(\rho^a, \rho^b, \rho^c)$ for a given $(a,b,c)$ is said to achieve a multiplexing gain $r$ and diversity gain $d$ if $$\lim_{\rho\rightarrow\infty}\frac{R(\rho^a, \rho^b, \rho^c)}{\log\rho} = r,\;\;\lim_{\rho\rightarrow\infty}\frac{P_e(\rho^a, \rho^b, \rho^c)}{\log\rho} = -d.$$
For each multiplexing gain $r$, the supremum $d(r)$ of diversity gains achievable over all families of codes is called the diversity-multiplexing tradeoff (DMT) of the half-duplex $(a,b,c)$-relay channel. 

In the setup of Figure~\ref{subfig:parallel}, the source communicates to the destination through two half-duplex parallel relays, i.e. there is no broadcasting from the source and no superposition at the destination. Here, we only focus on the case where all channels have the same average SNR, which turns out to be sufficient for demonstrating the necessity of a dynamic QMF strategy.

\section{Half-Duplex $(a,b,c)$-relay channel}\label{sec:relay}
\squeezeup \subsection{The Full-Duplex DMT}
\squeezeup We first derive the generalized diversity-multiplexing trade-off of the full-duplex $(a,b,c)$-relay channel. This serves as an upper bound for the optimal DMT of the corresponding half-duplex channel.

\begin{lem} \label{lem:fullduplex}
The diversity-multiplexing tradeoff of the full-duplex $(a,b,c)$-relay channel is given by $$d_{f.d.}(r)=\left(\min(a,b)-r\right)^{+}+(c-r)^{+}.$$
\end{lem}
\emph{Proof of Lemma~\ref{lem:fullduplex}:} \cite{AveDigTse} shows that the capacity of the full-duplex relay channel $C_{f.d}$ is bounded as $C_u-\kappa_1\leq C_{f.d.}\leq C_u+\kappa_2$ where both $\kappa_1$ and $\kappa_2$ are constants independent of SNR and 
\begin{IEEEeqnarray}{rCl}
C_u & = & \min\left\{\log(1+|h_{sr}|^2\rho^a + |h_{sd}|^2\rho^c)\right.,\nonumber\\
& &\qquad\quad \left.\log(1+|h_{rd}|^2\rho^b + |h_{sd}|^2\rho^c)\right\}.\nonumber
%& \doteq & \min\left\{\max(\alpha,\gamma),\max(\beta,\gamma)\right\}\log\rho
\end{IEEEeqnarray}
This implies that in the high-SNR limit the outage event $\{C_{f.d}\leq r\log \rho\}$ is equivalent to 
 $$\mathcal{O}(r) = \left\{\min\left(\max(\alpha,\gamma),\max(\beta,\gamma)\right) \leq r\right\}$$
where $\alpha= \lim_{\rho\rightarrow\infty}\frac{\log(1+\left|h_{sr}\right|^2\rho^a)}{\log\rho}$ is the exponential order of $\left|h_{sr}\right|^2$. Similarly $\beta,\gamma$ are the exponential orders of $|h_{rd}|^2$ and $|h_{sd}|^2$ respectively. The outage probability is given by $P_{\mathcal{O}(r)}\doteq \rho^{-d_{f.d.}(r)}$ where $\doteq$ implies that
$\frac{\log P_{\mathcal{O}(r)}}{\log\rho}\rightarrow -d_{f.d.}(r)$
asymptotically in $\rho$ and from \cite{ZheTse},
\begin{IEEEeqnarray}{l}
d_{f.d.}(r) \;\;=\;\; \min\;  a+b+c-\alpha-\beta-\gamma \nonumber\\
    \textrm{s.t. }\quad\min\left(\max(\alpha,\gamma),\max(\beta,\gamma)\right) \leq r,\nonumber\\
  \quad\quad 0\leq\alpha\leq a,\;0\leq \beta\leq b,\;0\leq \gamma\leq c.\label{eq:optprob}
\end{IEEEeqnarray}
It is straightforward to show that this minimization yields the result in the lemma.\hfill$\QED$
%The minimization problem is straightforward to solve. Assume without loss of generality that $a\leq b$: when $r\leq a$, the minimum is achieved by $?$; when $r>a$ the minimum is achieved by $?$. This completes the proof of the lemma.
\subsection{QMF with a fixed schedule}

We now investigate a static QMF strategy: here, the relay listens for half of the total duration for communication, then quantizes its received signal at the noise level and maps it to a random codeword, and transmits it in the second half. This fixed TX-RX schedule is independent of the channel realizations. The strategy uses only CSIR at the relay to determine the noise level for quantization.

\begin{lem}\label{lem:statqmf}
The DMT achieved by static QMF on the half-duplex \((a,b,c)\)-relay channel is given by $d_{QMF}(r)=$
\[
\begin{cases}
	\left(\min(a,b)-r\right)^{+}+(c-r)^{+} & \text{if } c > \min(a,b) \\ 
	\left(\min(a,b)+c-2r\right)^{+} & \text{if } c \leq \min(a,b).
	\end{cases}
\]
\end{lem}

Comparing Lemma~\ref{lem:fullduplex} and Lemma~\ref{lem:statqmf}, we immediately have the following corollary.
\begin{cor}\label{cor:sqmf}
Static QMF is optimal and achieves the full duplex DMT in the half-duplex $(a,b,c)$ relay channel when
\begin{itemize} 
\item $c\geq\min(a,b)$,
\item $c<\min(a,b)$ and $r\leq c$.
\end{itemize}
\end{cor}

This result shows that the half-duplex constraint does not manifest in the optimal DMT as long as $c\geq\min(a,b)$ and can be achieved with a fixed schedule, extending the result for $(a,b,c)=(1,1,1)$ from \cite{PawAveTse}. A similar conclusion holds for $c<\min(a,b)$ but only when $r\leq c$.

\begin{figure*}[!t]
\normalsize
\begin{IEEEeqnarray}{rCl}
\label{eq:hdcutset}C_{h.d.} &= & \min\left\{t\log(1+|h_{sr}|^2\rho^a +|h_{sd}|^2\rho^c)+(1-t)\log(1+|h_{sd}|^2\rho^c),\right.\nonumber\\ 
&&\left.\qquad\qquad t\log(1+|h_{sd}|^2\rho^c)+(1-t)\log(1+|h_{sd}|^2\rho^c +|h_{rd}|^2\rho^b)\right\}
%& \rightarrow & \min\left\{t\max(\alpha,\gamma) +(1-t)\gamma,t\gamma+(1-t)\max(\beta,\gamma)\right\}\log\rho &\quad = r_{QMF}\log\rho
\end{IEEEeqnarray}\squeezeup\squeezeup\squeezeup\squeezeup
\end{figure*}

\emph{Proof of Lemma~\ref{lem:statqmf}:} The rate of QMF $R_{QMF}$ in half-duplex networks with a {\em fixed} RX-TX schedule for the relays is lower bounded in \cite[Section~VIII-C]{AveDigTse}. If the relay listens for a fraction $t$ of the total time and transmits for $1-t$, this lower bound is $R_{QMF}\geq C_{h.d}-\kappa$, where $\kappa$ is a constant independent of SNR and $C_{h.d}$ is defined in \eqref{eq:hdcutset}. (Note that $t=1/2$ for the strategy that we are considering here.) The strategy is in outage if $R_{QMF}\leq r\log\rho$, which in the high SNR limit is equivalent to $r_{h.d.}\leq r$, where %$r_{h.d.}=\lim_{\rho\rightarrow\infty}\frac{C_{h.d.}}{\log\rho}=$ 
\begin{equation}\label{eq:rhd}
r_{h.d.}=\min\left\{t\max(\alpha,\gamma) +(1-t)\gamma,t\gamma+(1-t)\max(\beta,\gamma)\right\}\end{equation}
Therefore, the outage probability is given asymptotically by $P_{\mathcal{O}(r)}\doteq\rho^{-d_{QMF}(r)}$, where
\begin{IEEEeqnarray}{C}
d_{QMF}(r) \;\;=\;\; \min\;  a+b+c-\alpha-\beta-\gamma \nonumber\\
    \textrm{s.t. }\quad r_{h.d.} \leq r,\; 0\leq\alpha\leq a,\;0\leq \beta\leq b,\;0\leq \gamma\leq c.\nonumber
\end{IEEEeqnarray}
Solving this minimization yields the result in Lemma~\ref{lem:statqmf}.\hfill$\QED$

\subsection{Dynamic Decode and Forward}
We focus on the remaining case $c<\min(a,b)$ in the rest of this work. We next establish the DMT achieved by dynamic decode and forward
(DDF). Here the relay node waits until it it is able to decode the
transmitted message from the source. It then re-encodes the message
via a randomly chosen Gaussian codebook and transmits it. The
destination node chooses the most likely message in the source
codebook given its observation. The fraction of time the relay listens is determined dynamically depending on the transmission rate and the instantaneous realization of the S-R link.

Following \cite{AzaGamSch}, the fraction of time the relay needs to listen to decode the message from the source is $t = \frac{r\log\rho}{\log(1+|h_{sr}|^2\rho^a)} \rightarrow \frac{r}{\alpha}$ asymptotically in $\rho$. Outage occurs if:
\begin{itemize}
\item $t =\frac{r}{\alpha}> 1$ and $\gamma < r$: In this case, the relay never gets a chance to transmit and the direct link is not strong enough to support the desired rate alone, or
\item $t = \frac{r}{\alpha}\leq 1$ and $t\gamma+(1-t)\max(\gamma,\beta) < r$: In this case, the relay decodes and transmits but the mutual information acquired over the S,R--D cut is not sufficient to support the desired rate.
\end{itemize}
As before, the DMT of this strategy is given by $d_{DDF}(r) = \min\; a+b+c-\alpha-\beta-\gamma$ given the system is in outage. The first outage event is not critical as it requires $\alpha<r$ and  $\gamma < r$ in which case $d_{DDF}(r)\geq d_{f.d.}(r)$. The second outage event yields the following domain for the minimization problem:
$$
 t\gamma+(1-t)\max(\gamma,\beta) < r,\frac{r}{a}\leq t\leq 1,\;0\leq \beta\leq b,\;0\leq \gamma\leq c.
$$
Solving it, we arrive at following lemma.

%For the first case, the optimization problem we need to consider is
%\begin{IEEEeqnarray}{C}
%d_{DDF}(r) \;\;=\;\; \min\;  a+b+c-\frac{r}{t}-\beta-\gamma \nonumber\\
%  \text{s.t. }\quad t>1,\;0\leq \beta\leq b,\;0\leq \gamma\leq c,\; \gamma<r.\nonumber
%\end{IEEEeqnarray}
%which results in a DMT $\geq d_{f.d.}(r).$ For the second case, the constraints in the optimization problem become:
%$$
% t\gamma+(1-t)\max(\gamma,\beta) < r,\frac{r}{a}\leq t\leq 1,\;0\leq \beta\leq b,\;0\leq \gamma\leq c.
%$$
%Solving this problem, we arrive at Lemma \ref{lem:ddf}.

\begin{lem}\label{lem:ddf}
The diversity-multiplexing tradeoff achieved by DDF on the half-duplex \((a,b,c)\)-relay channel when $c<\min(a,b)$ is given by $d_{DDF}(r)=$
$$
\begin{cases}
	\min(a,b)+c-2r & \text{if } 0 \leq r \leq \min\left(c,\frac{\max(a,b)}{2}\right), \\
	\min(a,b) - \frac{(\max(a,b)-c)r}{\max(a,b)-r} & \text{if } c < r < \frac{\max(a,b)}{2},\\
	\frac{ab}{r}-a-b+c&\text{if } r\geq \max\left(\frac{a}{2},\frac{b}{2}\right).
	\end{cases}
$$
\end{lem}
\subsection{Upper Bound on the DMT with global CSI} 
We next turn to proving upper bounds on the achievable DMT that are tighter than the full duplex upper bound. In this section, we bound the achievable DMT under the optimistic assumption that the relay not only knows its incoming channel state but all the channel states in the network and can optimize its TX-RX times accordingly (global CSI). This obviously upper bounds the achievable DMT when the relay only has CSIR. In the next section, we derive a tighter upper bound on the achievable DMT with only CSIR at the relay.    

In the current and the next section, we restrict our attention to the case when $a$ and $b$ are equal. Let $a=b=p$.\footnote{Future work will focus on extending our results to $a\neq b$.}  Recall that we are considering $c<p$ since when $c\geq p$ we have shown in the earlier sections that static QMF is DMT optimal. The upper bound of the current section, establishes yet another regime where static QMF with half TX-half RX schedule achieves the optimal DMT. We show that when $r\geq \frac{p}{2}$, static QMF achieves the optimal DMT but falls short of achieving the full-duplex performance.

\begin{lem}\label{lem:dqmf}
The DMT of the half-duplex $(a,b,c)$-relay channel with global CSI $d_{G-CSI}(r)$ is upper bounded by
\begin{equation*}\label{eq:dqmf}
\min_{(\alpha,\beta,\gamma)\in\mathcal{O}(r)} a+b+c-\alpha-\beta-\gamma,
\end{equation*}
\begin{equation}\label{eq:globaloutage}
\mathcal{O}(r) = \left\{(\alpha,\beta,\gamma): \begin{array}{cc}\frac{\alpha\beta-\gamma^2}{\alpha+\beta-2\gamma} \leq r,\\ \gamma<\min(\alpha,\beta),\end{array} \begin{array}{cc}0\leq\alpha\leq a, \\ 0\leq\beta\leq b, \\ 0\leq\gamma\leq c\end{array}\right\}.\end{equation}
\end{lem}
\emph{Proof of Lemma~\ref{lem:dqmf}:} In \cite[Section VI]{OzgDig2}, an upper bound on the capacity of half-duplex relay networks when all channels are globally known is derived, which for the setup of Figure \ref{subfig:abc_relay} becomes $C\leq \max_{t(\alpha,\beta,\gamma)}C_{h.d.}+G$ where $G$ is a constant independent of SNR and $C_{h.d.}$ is given in \eqref{eq:hdcutset}. $t$ can be optimized as a function of {\em all} channel realizations. This yields the following upper bound on the DMT: 
\begin{IEEEeqnarray}{rCl}
d_{G-CSI}(r) & \leq & \min_{(\alpha,\beta,\gamma)\in\mathcal{O}(r)}\;\max_{t(\alpha,\beta,\gamma)} a+b+c-\alpha-\beta-\gamma,\nonumber%\label{eq:globalopt}
\end{IEEEeqnarray}
\begin{IEEEeqnarray*}{l}
\mathcal{O}(r)=\left\{0\leq\alpha\leq a,\;0\leq\beta\leq b,\;0\leq\gamma\leq c,\; r_{h.d.}\leq r \right\},
\end{IEEEeqnarray*}
%\begin{IEEEeqnarray}{rCl}
%d(r) & = & \min_{(\alpha,\beta,\gamma)\in\mathcal{O}(r)}\;\max_{t\in\mathcal{O}(r,\alpha,\beta,\gamma)} s(\alpha,\beta,\gamma),\label{eq:globalopt}
%\end{IEEEeqnarray}
%\begin{IEEEeqnarray*}{l}
%\mathcal{O}(r)=\left\{(\alpha,\beta,\gamma):0\leq\alpha\leq a,\;0\leq\beta\leq b,\;0\leq\gamma\leq c\right\},\\
%\mathcal{O}(r,\alpha,\beta,\gamma)=\left\{t:0\leq t\leq 1,\; r_{h.d.}\leq r\right\},
%\end{IEEEeqnarray*}
where $r_{h.d.}$ has been defined in \eqref{eq:rhd}. If $\gamma\geq\min(\alpha,\beta)$, we get a diversity $\geq d_{f.d.}(r)$ and the bound is no tighter than the full-duplex upper bound. So, we concentrate on $\gamma<\min(\alpha,\beta)$. It is easy to see that the optimal choice of $t$ is obtained by equating the two terms in $r_{h.d.}$ and  when $\gamma<\min(\alpha,\beta)$ $t=\frac{\beta-\gamma}{\alpha+\beta-2\gamma}$. Substituting this in $r_{h.d.}$ gives \eqref{eq:globaloutage}. This completes the proof of the lemma.\hfill $\QED$

 %The next two results, together with Corollary \ref{cor:sqmf}, generalize the earlier special results in \cite{PawAveTse} and \cite{GunKhoGolPoo}. We can also understand now when the half-duplex constraint and the receiver-CSI constraint have an effect on the DMT.
\begin{lem}\label{lem:abstatqmf}
When $c<a=b=p$, static QMF (with equal listening and transmit times) is optimal for $r\geq \frac{p}{2}$ on the half-duplex $(a,b,c)$-relay channel.
\end{lem}
\emph{Proof of Lemma~\ref{lem:abstatqmf}:} A critical outage event for the static QMF protocol for $r\geq\frac{p}{2}$ when $a=b=p$ is $(\alpha,\beta,\gamma)=\left(p,p,2r-p\right)$. It can be verified that $(p,p,2r-p)\in \mathcal{O}(r)$ in \eqref{eq:globaloutage}. Therefore, $d_{G-CSI}(r)\leq p+c-2r$ which is achieved by static QMF.\hfill$\QED$

\subsection{Upper Bound on the DMT with local CSIR}
\vspace{-1mm} We next establish an upper bound on the optimal DMT when the relay has only CSIR. This upper bound shows that DDF is optimal under CSIR in the range $c<r<\frac{p}{2}.$ To the best of our knowledge, upper bounds on the optimal DMT under limited CSI do not appear in the literature.

\begin{thm}\label{thm:ddfreceivecsi}
When $a=b=p$ and $c < r < \frac{p}{2}$, the optimal DMT of the half-duplex $(a,b,c)$-relay channel with CSIR is attained by DDF.
\end{thm}
\emph{Proof of Theorem~\ref{thm:ddfreceivecsi}:} The upper bound in \cite[Section VI]{OzgDig2} can be adopted to the case of limited CSI, when the TX-RX times are allowed to depend only on some of the channel realizations. For the single relay channel with CSIR at the relay, it gives $C\leq \max_{t(\alpha)}C_{h.d.}+G$ where $t(\alpha)$ is any function of $\alpha$ taking values in $[0,1]$. %The constraint of receiver-CSI is manifested in the order of the minimizations and maximizations in the optimization problem that we need to consider:
This provides the following upper bound on the DMT:
\begin{equation}\label{eq:localopt}
d_{L-CSI}(r) \leq \min_{\alpha\in[0,p]}\;\max_{t\in[0,1]}\;\min_{(\beta,\gamma)\in\mathcal{O}(r,\alpha,t)} s(\alpha,\beta,\gamma)
\end{equation} where 
$$s(\alpha,\beta,\gamma)=a+b+c-\alpha-\beta-\gamma,$$
$$\mathcal{O}(r,\alpha,t)=\left\{(\beta,\gamma): 0\leq\beta\leq p, \; 0\leq\gamma\leq c,\; r_{h.d.}\leq r\right\}.$$
The order of optimizations in \eqref{eq:localopt} means that nature chooses some $\alpha$ which we can observe and optimize $t$ accordingly (due to CSIR); however, nature gets a second round in which it can make adversarial choices for $(\beta,\gamma)$ depending on $\alpha$ and $t$. In other words, the RX time $t(\alpha)$ chosen by the relay should work equally well for all possible realizations of $(\beta,\gamma)$. This creates the following tension: if $t$ is chosen very small, so that the relay cannot decode the source message, the communication can be in outage if the S--D link turns out to be weak,  in which case we may not be able to convey sufficient mutual information over the S-R,D cut; whereas if we choose $t$ to be very large, so that the relay is left with little time to transmit, the R-D link can take on values that make the S,R--D cut sufficiently weak so as to cause outage. This intuition is formalized in the following analysis.
%(For a static protocol, the maximization over $t$ in \eqref{eq:localopt} would be outermost, while for a protocol with global-CSI, it would be innermost.) 
We fix $\alpha = p$ to get:
$$
d_{L-CSI}(r) \leq \max_{t\in[0,1]}\;\min_{(\beta,\gamma)\in\mathcal{O}(r,p,t))} s(p,\beta,\gamma).
$$
\begin{itemize}
\item If $t < r/p$, $(\beta,\gamma)=(p,0)$ is a feasible point in the above minimization problem. Hence $d(r)\leq c.$
\item If $t\geq r/p$, $\beta=\min\left(\frac{r-tc}{1-t},p\right)$, $\gamma=c$ is feasible. Hence 
$$
d(r) \leq p -\min\left(\frac{r-tc}{1-t},p\right) \leq p - \frac{(p-c)r}{p-r}\quad\text{ if } r>c.
$$
\end{itemize}
This shows that $d_{L-CSI}(r)\leq \max\left(c, p - \frac{(p-c)r}{p-r}\right)$. Since DDF achieves $p - \frac{(p-c)r}{p-r}$ which is equal to $\max\left(c, p - \frac{(p-c)r}{p-r}\right)$ for $c<r<\frac{p}{2}$, this proves that DDF is optimal for $c<r<\frac{p}{2}$  when the relay has only CSIR.\hfill$\QED$

%Table \ref{table:dmt11n} summarizes the results for the half-duplex $(a,b,c)$-relay channel for the case $(a,b,c)=(1,1,\eta)$, $\eta\leq 1$. The two extremes in \cite{PawAveTse} and \cite{GunKhoGolPoo} correspond to $\eta=1$ and $\eta=0$ respectively.

%\begin{table*}
%\renewcommand{\arraystretch}{1.3}
%\caption{DMT for half-duplex $(1,1,\eta)$ relay channel with receiver CSI ($\eta\leq 1$)}
%\label{table:dmt11n}
%\centering
%\begin{tabular}{c|c|c|c|c|c}
%\hline
%\bfseries &Multiplexing gains& Optimal DMT & Optimal protocol & Comparison with full-duplex & receiver-CSI vs global CSI\\ \hline\hline
%$\eta > 1/2$ & $0\leq r\leq\frac{1+\eta}{2}$ & $1+\eta -2r$ & Static QMF & & not restrictive\\ \hhline{----~-}  
%\multirow{3}{*}{$\eta \leq 1/2$} & $0\leq r\leq\eta$ & $1+\eta -2r$ & Static QMF, DDF & = FD DMT for $r<\eta$ & not restrictive\\
%& $\eta \leq r \leq \frac{1}{2}$ & $1-\frac{r(1-\eta)}{1-r}$ & DDF & $\neq$ FD DMT for $r\geq\eta$ & maybe restrictive\\
%& $\frac{1}{2} \leq r \leq \frac{1+\eta}{2}$ & $1+\eta -2r$ & Static QMF & & not restrictive\\ \hline
%\end{tabular}
%\end{table*}

\section{Half-Duplex Parallel Relay channel}\label{sec:parallel}
A natural generalization of these two strategies, DDF and static QMF discussed in the earlier sections, is dynamic QMF where the relay listens for a fraction of time determined by its CSIR that is not necessarily long enough to allow decoding. It then quantizes maps and forwards the received signal as in the static QMF. This generalization was not necessary in the earlier sections, but through the parallel relay network in Fig~\ref{subfig:parallel}, \cite{OzgDig} shows that it is needed to achieve the optimal trade-off in more general networks. The difficulty in applying this strategy is in identifying the optimal (dynamic) choice of the listening times, which was left open in \cite{OzgDig}.

In this paper, we identify  a simple optimal (dynamic) schedule for the relays. In the context of the single relay channel of the earlier sections, say when $a=b=1$ and $c<1$ this dynamic choice corresponds to $t=1-\alpha(1-r)$ instead of $t=r/\alpha$ in DDF. While $t=r/\alpha$ is chosen to ensure that the relay can decode the transmitted message, the choice $t=1-\alpha(1-r)$ is motivated by the need to balance the multiplexing gain achieved over the two cuts of the network dynamically, based only on the observation of $\alpha$.
Note that when $\alpha$ is large $t$ is small, and the strategy allocates more time to the second stage which helps in case the second stage turns out to be weak. When $\alpha$ is small, the relay allocates more time to listen. Indeed, it can be readily observed that when $\frac{r}{1-r}\leq\alpha\leq 1$ (this is the range of $\alpha$'s where DDF is not in outage)  $1-\alpha(1-r)> r/\alpha$, the relay can always decode and it turns out that the two strategies are equivalent for the single relay channel. However, as we show next this is not the case for the parallel relay channel. While dynamic QMF reaches the best achievable DMT with global CSI, DDF (and also static QMF) fails to do so.
%The expression for DMT in Lemma \ref{lem:dqmf3} can be verified by substituting for $t_1$ and $t_2$ in the outage expression and considering the three cases: $\alpha+\beta\leq\frac{1}{1-r},\gamma+\delta\leq\frac{1}{1-r}$, $\alpha+\beta\leq\frac{1}{1-r},\gamma+\delta>\frac{1}{1-r}$ and $\alpha+\beta>\frac{1}{1-r},\gamma+\delta>\frac{1}{1-r}$.

\begin{thm}\label{thm:parallelmain}
The optimal DMT of the half-duplex parallel-relay channel in Figure~\ref{subfig:parallel} with CSIR is given by 
$$
d(r) = 
	\begin{cases}
	2-\frac{r}{1-r},& 0\leq r\leq \frac{1}{2}\\
	2(1-r), & \frac{1}{2}\leq r\leq 1
	\end{cases}
$$
When $0\leq r<1/2$, the optimal DMT is achieved by a dynamic QMF strategy with switching times chosen as $t_1 = 1-\alpha(1-r)$ and $t_2 = 1-\gamma(1-r)$, where $\alpha$ and $\gamma$ are the exponential orders of the channels in the first stage, $h_{sr_1}$ and $h_{sr_2}$ respectively.
When $1/2\leq r \leq 1$, the optimal DMT is achieved by a static QMF strategy with equal times for transmitting and receiving.
\end{thm}
\emph{Main Idea of the Proof of Theorem~\ref{thm:parallelmain}:}
\begin{itemize}
%\item $0\leq r\leq\frac{1}{2}$: The critical outage point of the dynamic QMF strategy in Lemma \ref{lem:dqmf3} $(\alpha,\beta,\gamma,\delta)=\left(1,\frac{r}{1-r},1,0 \right)$ is a feasible point of \eqref{eq:dqmf2} for $0\leq r\leq\frac{1}{2}$, so Lemma \ref{lem:dqmf2} gives us the desired result.
\item $0\leq r\leq\frac{1}{2}$: As in Lemma \ref{lem:dqmf}, an upper bound can be obtained by assuming that the relays have global CSI. 
%In this case, the optimal listening times for the 2 relays are $t_1=\beta/(\alpha+\beta)$, $t_2=\delta/(\gamma+\delta)$. 
The dynamic QMF strategy described above meets this upper bound. The DMT achieved by static QMF (equal times for listening and transmitting) is $2-2r$ and that by DDF can be shown to be $\leq 2-2r$ \cite{OzgDig}.%, so for $0\leq r\leq 1/2$, neither DDF nor static QMF is able to achieve the optimal DMT.
%which gives us an upper bound:
%\begin{equation}\label{eq:dqmf2}
%$d_{G-CSI}(r)\leq\min_{(\alpha,\beta,\gamma,\delta)\in\mathcal{O}(r)} 4-\alpha-\beta-\gamma-\delta,$ where $\mathcal{O}(r) =$
%\end{equation}
%\begin{equation}\label{eq:dqmf2}
%$$\left\{(\alpha,\beta,\gamma,\delta): \frac{\alpha\beta}{\alpha+\beta}+\frac{\gamma\delta}{\gamma+\delta} \leq r,\; 0\leq\alpha,\beta,\gamma,\delta\leq 1 \right\}.$$%\end{equation} 
%A critical outage point of the dynamic QMF strategy in Lemma \ref{lem:dqmf3} is $\left(1,\frac{r}{1-r},1,0 \right)$ which is feasible in this upper bound for $0\leq r\leq\frac{1}{2}$.
\item $\frac{1}{2}\leq r\leq 1$: As in Theorem~\ref{thm:ddfreceivecsi}, we can write down an upper bound to the DMT under CSIR. This upper bound can be further upper bounded by fixing $\alpha=\gamma=1$. We can then show that whatever switching time the relays decide upon, adversarial choices for the channel realizations in the second stage ensure that $d(r)\leq 2-2r,$ which is achieved by static QMF \cite{OzgDig}.\hfill$\QED$
%\begin{equation}\label{eq:parallelreceiveCSI}
%$$d(r)=\min_{\substack{(\alpha,\gamma)\in\\\mathcal{O}(r)}}\;\max_{\substack{(t_1,t_2)\in\\\mathcal{O}(r,\alpha,\gamma)}}\;\min_{\substack{(\beta,\delta)\in\\ \mathcal{O}(r,\alpha,\gamma,t_1,t_2)}} 4-\alpha-\beta-\gamma-\delta

%where the outage regions are defined in a similar manner to Theorem \ref{thm:ddfreceivecsi} and $t_1=f(r,\alpha)$, $t_2=f(r,\gamma)$ where $f(\cdot,\cdot)\in[0,1]$. Note that both relays use the same function $f$ by symmetry. %Setting $\alpha=1$ and $\gamma=1$:
%$$
%d(r) \leq \max_{\mathcal{O}(r,1,1)}\;\min_{\mathcal{O}(r,1,1,f(r,1),f(r,1))} 2-\beta-\delta
%$$
%We set $\alpha=\gamma=1$ and consider 3 cases separately: $\{f(r,1)\leq\frac{r}{2}\}$, $\{\frac{r}{2} < f(r,1) \leq \frac{1}{2}\}$ and $\{f(r,1)>\frac{1}{2}\}$. Taking the maximum over the 3 cases we obtain $d(r)\leq 2-2r.$
%\begin{itemize}
%\item If $f(r,1)\leq\frac{r}{2}$, $\beta=\delta=1$ is feasible $\Rightarrow d(r)=0$.
%\item If $f(r,1)>\frac{1}{2}$, $\beta=\delta=\frac{r}{2(1-f(r,1))}$ is feasible $\Rightarrow d(r) \leq 2 - \frac{r}{1-f(r,1)} \leq 2-2r.$
%\item If $\frac{r}{2} < f(r,1) \leq \frac{1}{2}$, $\beta=\frac{r-f(r,1)}{1-f(r,1)}$, $\delta=1$ is feasible. So $d(r) \leq 1 - \frac{r-f(r,1)}{1-f(r,1)} \leq 1 - \frac{r-\frac{1}{2}}{1-\frac{1}{2}}=2-2r.$
%\end{itemize}
%So $d(r)\leq\max(0,2-2r,2-2r)= 2-2r$. 
\end{itemize}

%\section{Conclusion}

%Table \ref{table:dmt11n} summarizes the results for the half-duplex $(a,b,c)$-relay channel for the special case $(a,b,c)=(1,1,\eta)$. Results for $\eta=1$ in \cite{PawAveTse} and $\eta=0$ in \cite{GunKhoGolPoo} can now be put into perspective. For large values of $\eta$, i.e. when the strength of the direct channel (as measured by proximity gain) is comparable to the source-relay and relay-destination links, requiring the relay to decode the message is suboptimal. When the direct channel is significantly weaker $(\eta\leq 1/2)$, optimal DMT at low multiplexing gains $(0\leq r\leq 1/2)$ is obtained by dynamically decoding the message at the relay. At very low $(0\leq r\leq\eta)$ and high multiplexing gains $(r\geq 1/2)$, static QMF is optimal. When the direct channel is absent, all multiplexing gains fall in the range where DDF is optimal. 

%Table \ref{table:dmt11n} also highlights when the optimal DMT is equal to the full-duplex DMT and when the receiver-CSI assumption is not restrictive. We characterize the optimal DMT under the constraint of receiver-CSI which is a more practical assumption than global CSI. 

%We also answer the question of whether it is sufficient to consider static QMF and DDF by analyzing the half-duplex parallel relay channel where a dynamic QMF protocol is necessary to achieve the optimal DMT.

%% Appendix:
%% If needed a single appendix is created by
%\appendix
%\appendices,\section{},\section{}

%% Use \section* for acknowledgement
%\squeezeup
%\section*{Acknowledgment}
%R. Kolte is supported by a Stanford Graduate Fellowship. 

\bibliographystyle{IEEEtran}
\bibliography{IEEEabrv,dmt}

\begin{thebibliography}{10}
\providecommand{\url}[1]{#1}
\csname url@samestyle\endcsname
\providecommand{\newblock}{\relax}
\providecommand{\bibinfo}[2]{#2}
\providecommand{\BIBentrySTDinterwordspacing}{\spaceskip=0pt\relax}
\providecommand{\BIBentryALTinterwordstretchfactor}{4}
\providecommand{\BIBentryALTinterwordspacing}{\spaceskip=\fontdimen2\font plus
\BIBentryALTinterwordstretchfactor\fontdimen3\font minus
  \fontdimen4\font\relax}
\providecommand{\BIBforeignlanguage}[2]{{%
\expandafter\ifx\csname l@#1\endcsname\relax
\typeout{** WARNING: IEEEtran.bst: No hyphenation pattern has been}%
\typeout{** loaded for the language `#1'. Using the pattern for}%
\typeout{** the default language instead.}%
\else
\language=\csname l@#1\endcsname
\fi
#2}}
\providecommand{\BIBdecl}{\relax}
\BIBdecl

\bibitem{PawAveTse}
S.~Pawar, A.~Avestimehr, and D.~Tse, ``Diversity-multiplexing tradeoff of the
  half-duplex relay channel,'' in \emph{Allerton Conf. on Communication,
  Control, and Computing}, sep. 2008, pp. 27 --33.

\bibitem{GunKhoGolPoo}
D.~Gunduz, M.~Khojastepour, A.~Goldsmith, and H.~Poor, ``Multi-hop mimo relay
  networks: diversity-multiplexing trade-off analysis,'' \emph{{IEEE} Trans.
  Wireless Commun.}, vol.~9, no.~5, pp. 1738 --1747, may 2010.

\bibitem{ZheTse}
L.~Zheng and D.~Tse, ``Diversity and multiplexing: a fundamental tradeoff in
  multiple-antenna channels,'' \emph{{IEEE} Trans. Inf. Theory}, vol.~49,
  no.~5, pp. 1073 -- 1096, may 2003.

\bibitem{LanTseWor}
J.~Laneman, D.~Tse, and G.~Wornell, ``Cooperative diversity in wireless
  networks: Efficient protocols and outage behavior,'' \emph{{IEEE} Trans. Inf.
  Theory}, vol.~50, no.~12, pp. 3062 -- 3080, dec. 2004.

\bibitem{SenErkAaz}
A.~Sendonaris, E.~Erkip, and B.~Aazhang, ``User cooperation diversity. part i.
  system description,'' \emph{{IEEE} Trans. Commun.}, vol.~51, no.~11, pp. 1927
  -- 1938, nov. 2003.

\bibitem{AzaGamSch}
K.~Azarian, H.~El~Gamal, and P.~Schniter, ``On the achievable
  diversity-multiplexing tradeoff in half-duplex cooperative channels,''
  \emph{{IEEE} Trans. Inf. Theory}, vol.~51, no.~12, pp. 4152 --4172, dec.
  2005.

\bibitem{AveDigTse}
A.~Avestimehr, S.~Diggavi, and D.~Tse, ``Wireless network information flow: A
  deterministic approach,'' \emph{{IEEE} Trans. Inf. Theory}, vol.~57, no.~4,
  pp. 1872 --1905, apr. 2011.

\bibitem{YukErk}
M.~Yuksel and E.~Erkip, ``Multiple-antenna cooperative wireless systems: A
  diversity-multiplexing tradeoff perspective,'' \emph{{IEEE} Trans. Inf.
  Theory}, vol.~53, no.~10, pp. 3371 --3393, oct. 2007.

\bibitem{EtkTseWan}
R.~Etkin, D.~Tse, and H.~Wang, ``Gaussian interference channel capacity to
  within one bit,'' \emph{{IEEE} Trans. Inf. Theory}, vol.~54, no.~12, pp. 5534
  --5562, dec. 2008.

\bibitem{OzgJohTseLev}
A.~Ozgur, R.~Johari, D.~Tse, and O.~Leveque, ``Information-theoretic operating
  regimes of large wireless networks,'' \emph{{IEEE} Trans. Inf. Theory},
  vol.~56, no.~1, pp. 427 --437, jan. 2010.

\bibitem{NagPawTseNik}
V.~Nagpal, S.~Pawar, D.~Tse, and B.~Nikolic, ``Cooperative multiplexing in the
  multiple antenna half duplex relay channel,'' in \emph{IEEE Int. Symp. Inf.
  Theory}, 28 jun.-3 jul. 2009, pp. 1438 --1442.

\bibitem{KarVar}
S.~Karmakar and M.~Varanasi, ``The diversity-multiplexing tradeoff of the mimo
  half-duplex relay channel,'' \emph{{IEEE} Trans. Inf. Theory}, vol.~58,
  no.~12, pp. 7168--7187, 2012.

\bibitem{PraVar}
N.~Prasad and M.~Varanasi, ``High performance static and dynamic cooperative
  communication protocols for the half duplex fading relay channel,'' in
  \emph{IEEE GLOBECOM '06}, 2006, pp. 1--5.

\bibitem{OzgDig}
A.~Ozgur and S.~Diggavi, ``Dynamic qmf for half-duplex relay networks,'' in
  \emph{IEEE Int. Symp. Inf. Theory}, jul. 2012, pp. 413 --417.

\bibitem{OzgDig2}
------, ``Approximately achieving gaussian relay network capacity with lattice
  codes,'' in \emph{IEEE Int. Symp. Inf. Theory}, jun. 2010, pp. 669 --673.

\end{thebibliography}

\end{document}